\newcommand{\be}{\begin{equation}}\newcommand{\ee}{\end{equation}}
\newcommand{\bea}{\begin{eqnarray}}\newcommand{\eea}{\end{eqnarray}}
\newcommand{\nn}{\nonumber\\[6pt]}
\newcommand{\p}[1]{(\ref{#1})}

\newcommand{\bD}{\overline D}

\newcommand{\bpsi}{{\bar\psi}}

\newcommand{\blam}{{\bar\lambda}}
\newcommand{\bLam}{{\overline\Lambda}}

\documentclass[12pt]{article}
\usepackage{amscd,amsmath,amssymb}

\begin{document}

\thispagestyle{empty}
\vspace{2cm}
\begin{flushright}
\end{flushright}
\begin{center}
{\Large\bf Geometry of $N=4, d=1$ nonlinear supermultiplet}
\end{center}
\vspace{1cm}

\begin{center}
{\large\bf S. Bellucci${}^{a}$, S. Krivonos${}^{b}$ }
\end{center}

\begin{center}
${}^a$ {\it INFN-Laboratori Nazionali di Frascati,
Via E. Fermi 40, 00044 Frascati, Italy}

\vspace{0.2cm}

${}^b$ {\it Bogoliubov  Laboratory of Theoretical Physics, JINR, 141980 Dubna,
Russia}

\vspace{1cm}
bellucci@lnf.infn.it, krivonos@theor.jinr.ru

\end{center}
\vspace{2cm}

\begin{abstract}
We construct the general action for $N=4, d=1$ nonlinear supermultiplet including
the most general interaction terms which depend on the arbitrary function $h$ obeying the Laplace
equation on $S^3$.  We find the bosonic field $B$ which depends on the components of nonlinear supermultiplet and
transforms as a full time derivative under $N=4$ supersymmetry. The most general interaction is generated just
by a Fayet-Iliopoulos term built from this auxiliary component.

Being transformed through a full time derivative under $N=4, d=1$ supersymmetry, this auxiliary component $B$
may be dualized into a fourth scalar field giving rise to a four dimensional $N=4, d=1$ sigma-model.
We analyzed the geometry in the bosonic sector and find that it is not a hyper-K\"ahler one.
With a particular choice of the target space metric $g$ the geometry in the bosonic
sector coincides with the one which appears in heterotic $(4,0)$ sigma-model in $d=2$.
\end{abstract}

\newpage
\setcounter{page}{1}
\section{Introduction}
The one dimensional theories (mechanics) with extended supersymmetries possess a number of specific features which make
them selected among their higher-dimensional counterparts. In this respect the existence of nonlinear off-shell
supermultiplets in $d=1$ is an impressive example. The simplest examples of such multiplets are nonlinear ${\bf (3,4,1)}$
\cite{{il},{ikl}} and nonlinear ${\bf (2,4,2)}$ \cite{ikl} supermultiplets. These nonlinear supermultiplets have the same
component contents as their linear analogs while the transformation properties of their components are highly nonlinear
under supersymmetry. As a result, the geometry of the bosonic target space of the sigma models constructed from nonlinear
supermultiplets should be rather different from those in the linear cases. The supersymmetric mechanics constructed with
nonlinear ${\bf (2,4,2)}$ supermultiplet has been considered in details in \cite{n242}, while no such an exhaustive study
was undertaken so far for nonlinear ${\bf (3,4,1)}$ supermultiplet. The basic aim of this paper is to give a detailed
description of the $N=4$ supersymmetric mechanics with nonlinear ${\bf (3,4,1)}$ supermultiplet. We construct the most general
sigma-model type action for this nonlinear supermultiplet and extend it by the most general Fayet-Iliopoulos (FI) term.
Being equipped with a general FI term we, in full analogy with linear ${\bf (3,4,1)}$ supermultiplet ,
perform the dualization of an auxiliary component into a fourth physical boson, thus finishing with nonlinear ${\bf (4,4,0)}$ supermultiplet.
This new multiplet belongs to a new class of nonlinear supermultiplets which contain a functional freedom
in the defining relations \cite{{ks},{buks},{di},{bn}}. We construct the sigma-model action for this supermultiplet and explicitly
demonstrate that the bosonic target space geometry is not a hyper-K\"ahler one.

\setcounter{equation}0
\section{N=4 nonlinear multiplet and its action}
The nonlinear $N=4$ supermultiplet is a $d=1$ analog of the $N=2, d=4$ nonlinear multiplet \cite{{nlin},{klr},{HKLR}}.
It can be described by 4 by 4 matrix variables $N^{ai}\; (i=1,2; a=1,2)$ obeying the constraints \cite{ikl}
\be\label{N}
N^{ia} N_{ia}=2, \qquad\qquad N^{a(i}D^jN^{k)}_a=0,\; N^{a(i}\bD{}^jN^{k)}_a=0,
\ee
where $N=4,d=1$ spinor derivatives are defined by
\be
D^i =\frac{\partial}{\partial\theta_i}+i\bar\theta{}^i\partial_t,\;
\bD_i= \frac{\partial}{\partial \bar\theta{}^i}+i\theta_i\partial_t, \qquad
\left\{D^i \bD_j\right\}=2i \delta^i_j \partial_t.
\ee
This very symmetric description of the nonlinear supermultiplet by matrix superfields $N^{ia}$ is not
very useful in practice. The preferable one is by the following representation of $N^{ia}$ \cite{ikl}:
\be\label{new}
N^{11}=\frac{e^{-\frac{i}{2}\Phi}}{\sqrt{1+\Lambda\bLam}}\Lambda,\;
N^{21}=\frac{e^{\frac{i}{2}\Phi}}{\sqrt{1+\Lambda\bLam}},\;
N^{12}=-\frac{e^{-\frac{i}{2}\Phi}}{\sqrt{1+\Lambda\bLam}},\;
N^{22}=\frac{e^{\frac{i}{2}\Phi}}{\sqrt{1+\Lambda\bLam}}\bLam\;.
\ee
The representation \p{new} solves the algebraic constraint in \p{N} in terms of independent $N=4$ superfields
$\phi, \Lambda,\bLam$ which, as a consequence of the differential constraints in \p{N}, should obey
\bea\label{con1}
D^1 \Lambda = -\Lambda D^2 \Lambda,\quad \bD_2 \Lambda =\Lambda \bD_1 \Lambda, &&
D^2 \bLam = \bLam  D^1 \bLam,\quad \bD_1 \bLam = -\bLam \bD_2\bLam ,\nn
iD^1\Phi=-D^2 \Lambda,\quad i\bD_1 \Phi = \bD_2 \bLam, && iD^2 \Phi= -D^1 \bLam,\quad i\bD_2 \Phi = \bD_1 \Lambda.
\eea
{}From \p{con1} one may immediately reveal the component field content of the nonlinear supermultiplet which includes
three physical $\phi, \lambda,\blam$ and one auxiliary $A$ bosonic fields and four fermions $\psi_a,\bpsi{}^a$
defined as
\bea\label{comp}
&&\phi=\Phi|,\quad \lambda=\Lambda|, \quad\blam=\bLam|,\qquad A =(D^1\bD_1 -\bD_1 D^1)\Phi|, \nn
&& \psi_1=\frac{1}{2} \bD_1 \Phi|,\quad \psi_2 =-\frac{1}{2}\bD_2\Phi|,\qquad \bpsi^1 =-\frac{1}{2}D^1 \Phi|, \quad\bpsi^2=\frac{1}{2}D^2\Phi|,
\eea
where $|$ as usual means $\theta_i=\bar\theta^j=0$. Under $N=4$ supersymmetry these components transform as follows:
\bea\label{N4}
&& \delta\phi= 2\left( \epsilon_1 \bpsi{}^1 -\epsilon_2\bpsi{}^2 -\bar\epsilon{}^1\psi_1+\bar\epsilon{}^2\psi_2\right), \;
\delta\lambda= -2i\left(\epsilon_2 - \epsilon_1\lambda\right)\bpsi{}^1+2i\left( \bar\epsilon{}^1 +\lambda\bar\epsilon{}^2\right)\psi_2\nn
&& \delta \psi_1 =-\frac{1}{2} \epsilon_1 \left( i\dot{\phi}+\frac{1}{2}A\right) -\frac{1}{2}\epsilon_2\left(
 2\dot{\blam} +4i \psi_1\bpsi{}^2+i\blam\dot{\phi} +\frac{1}{2}\blam A\right), \nn
&& \delta \psi_2 =\frac{1}{2} \epsilon_2 \left( i\dot{\phi}-\frac{1}{2}A\right) +\frac{1}{2}\epsilon_1\left(
 2\dot{\lambda} -4i \psi_2\bpsi{}^1-i\lambda\dot{\phi} +\frac{1}{2}\lambda A\right), \nn
&& \delta A=-4i\left(\epsilon_1 \dot{\bar\psi}^1+\epsilon_2 \dot{\bpsi}^2+\bar\epsilon{}^1 \dot{\psi_1}+\bar\epsilon^2 \dot{\psi_2}\right).
\eea
One may check that, despite the presence of nonlinear terms, the transformations \p{N4} perfectly close to span $N=4, d=1$ super Poincar\'e
algebra. Finally, thanks to manifest $N=4$ supersymmetry, the general off-shell action
\be\label{action1}
S=\int dt d^2\theta d^2 \bar\theta L(\Phi, \Lambda,\bLam),
\ee
where $L(\Phi, \Lambda,\bLam)$ is an arbitrary real function of the superfields $(\Phi, \Lambda,\bLam)$, is invariant under $N=4,d=1$
supersymmetry. Being rewritten in terms of the components \p{comp}, the action \p{action1} reads
\be\label{action2}
S=\int dt\; g(1+\lambda\blam)\left[ \left( \dot\phi +i \frac{ \dot\lambda \blam -\lambda \dot\blam }{1+\lambda\blam}\right)^2+
\frac{4 \dot\lambda\dot\blam}{(1+\lambda\blam)^2} +\frac{1}{4}{\tilde A}^2+
\mbox{  fermions} \right],
\ee
where the metric $g$ is defined through the prepotential $L(\phi, \lambda,\blam)$ entering \p{action1} as
\be\label{g}
g= (1+\lambda\blam) L_{\lambda\blam}+L_{\phi\phi}+i\lambda L_{\lambda\phi}-i\blam L_{\blam\phi}
\ee
and the new auxiliary field $\tilde A$ reads
\be\label{tA}
\tilde A = A+2 \frac{d}{dt} \log (1+\lambda\blam).
\ee
For the sake of brevity we omitted in the action \p{action2} all fermionic terms, which may be easily reconstructed, if needed.

{}From now on, we have two options to go further. Firstly, one may immediately exclude the auxiliary field $\tilde A$ from the action \p{action2}.
The resulting action reads
\be\label{s1}
S=\int dt\; g(1+\lambda\blam)\left[ \left( \dot\phi +i \frac{ \dot\lambda \blam -\lambda \dot\blam }{1+\lambda\blam}\right)^2+
\frac{4 \dot\lambda\dot\blam}{(1+\lambda\blam)^2} +
\mbox{  fermions} \right].
\ee
It is clear that the terms in square brackets are just the sigma model action of the principal chiral field on $SU(2)$ \cite{ikl}. The curvature of
this three dimensional space is equal to 3/2. Thus, the full action \p{s1} describes the $N=4$ supersymmetric extension of the
particle moving on $S^3$ deformed by the conformal factor $g(1+\lambda\blam)$.

Alternatively, due to transformation properties of $A$ \p{N4}, one may dualize this components into a fourth scalar field as
\be\label{dual1}
\tilde A = 2\dot{y}.
\ee
Plugging \p{dual1} back in \p{action2} we get the four-dimensional $N=4$ sigma model with the following bosonic sector:
\be\label{s2}
S_{bosonic}=\int dt\; g(1+\lambda\blam)\left[ \left( \dot\phi +i \frac{ \dot\lambda \blam -\lambda \dot\blam }{1+\lambda\blam}\right)^2+
\frac{4 \dot\lambda\dot\blam}{(1+\lambda\blam)^2} +{\dot y}^2 \right].
\ee
In the simplest case when
\be
 g(1+\lambda\blam)=1 \qquad \Rightarrow \qquad L(\Phi,\Lambda,\bLam) = \log (1+\Lambda\bLam)
\ee
the action \p{s2} is just the action of $SU(2)\times U(1)$ sigma model.
Let us note that the corresponding target space is conformally flat.
Indeed, one may easily check that the action
\be\label{s2a}
S_1=\int dt\; e^{-y}\left[ \left( \dot\phi +i \frac{ \dot\lambda \blam -\lambda \dot\blam }{1+\lambda\blam}\right)^2+
\frac{4 \dot\lambda\dot\blam}{(1+\lambda\blam)^2} +{\dot y}^2 \right]
\ee
describes the particle in flat four-dimensional space. But the metric $g=\frac{e^{-y}}{(1+\lambda\blam)}$ is unreachable because
in our construction $g$ is a function depending on $(\phi, \lambda,\blam)$ only.

Another particular choice of the metric $g=e^{a\phi}$ produce the action
\be\label{s3a}
S_2=\int dt\; e^{a\phi}(1+\lambda\blam)\left[ \left( \dot\phi +i \frac{ \dot\lambda \blam -\lambda \dot\blam }{1+\lambda\blam}\right)^2+
\frac{4 \dot\lambda\dot\blam}{(1+\lambda\blam)^2} +{\dot y}^2 \right].
\ee
One may check that in this case all components of the curvature tensor are proportional to $(a^2 + 1)$. Therefore, within our
model with any choice of real function $g(\phi, \lambda,\blam)$ the flat metric is unreachable.

In the next section we will construct a more general action for the four-dimensional sigma model, but in any case there are
no chances for its bosonic manifold to be a hyper-K\"ahler one. Indeed, any more sophisticated four dimensional  sigma model action
will contain \p{s2a} as a particular solution. But within our approach the metric corresponding to the action \p{s2a} is not even conformally
a hyper-K\"ahler one. Thus the same will be true for any of its generalizations. Let us remind that in the linear case \cite{{ks},{buks}}
the analog of the action \p{s1} corresponds to a flat target space which is a rather particular case of a hyper-K\"ahler geometry.
Thanks to this fact there exists its generalization for a general case of a hyper-K\"ahler geometry with one isometry.

\setcounter{equation}0
\section{Towards N=4 four-dimensional sigma model}
In this section we will construct the potential terms for the nonlinear supermultiplet and show that there is
another dualization which gives rise to a four dimensional sigma-model with a geometry in the bosonic sector which depends on
two functions and is different from the hyper-K\"ahler one.

The simplest way to get the potential terms is to add the Fayet-Iliopoulos term to the action \p{action2}
\be\label{FI1}
{\tilde S} = S+ m\int dt {\tilde A}.
\ee
After excluding the auxiliary field $\tilde A$ in the action \p{FI1} we have the following potential term:
\be\label{pot1}
S_{pot}=-\int dt \frac{m^2}{g(1+\lambda\blam)}.
\ee
Clearly, this potential terms is not the most general one. In order to have  more possibilities for the interaction one
should construct a more general bosonic field $B$ which transforms as a full time derivative with respect to
$N=4$ supersymmetry \p{N4}. Starting from the most general Ansatz one may check that the most general real combination
of dimension $cm^{-1}$ that is composed of the nonlinear supermultiplet components and transforms as a total time
derivative has the following form:
\be\label{B}
B=fA +b\dot\lambda +{\bar b} \dot\blam +c\dot\phi+ a \left(\bpsi{}^1\psi_1 -\bpsi{}^2\psi_2\right)+
   a_1 \bpsi{}^2\psi_1+a_2\bpsi{}^1\psi_2,
\ee
where real dimensionless coefficients $(b, {\bar b},c,a,a_1,a_2)$ are expressed through the function $f$ depending
on $(\phi, \lambda,\blam)$ as
\bea\label{eq}
&& a= -8\frac{f_\phi}{1+\lambda\blam},\; a_1=-8if_{\blam}+8 \lambda\frac{f_\phi}{1+\lambda\blam},\;
a_2=8if_{\lambda}+8 \blam\frac{f_\phi}{1+\lambda\blam}, \\
&& b_\blam-{\bar b}_\lambda =-4i\frac{f_\phi}{1+\lambda\blam},\;c_\lambda-b_\phi=-2if_\lambda-4\blam\frac{f_\phi}{1+\lambda\blam},\;
c_\blam-{\bar b}_\phi=2if_\blam-4\lambda\frac{f_\phi}{1+\lambda\blam},\nonumber
\eea
while $f$ obeys the following equation:
\be\label{eq1}
f_{\phi\phi}+(1+\lambda\blam) f_{\lambda\blam}+i\lambda f_{\lambda\phi}-i\blam f_{\blam\phi}=0.
\ee
With all these equations satisfied, the new auxiliary component $B$ transforms under $N=4,d=1$ supersymmetry
as follows:
\be\label{trB}
\delta B =\frac{d}{dt} \left[ -4if \left( \epsilon_i \bpsi{}^i +\bar\epsilon{}^i\psi_i\right) + b\delta \lambda +
{\bar b} \delta \blam + c \delta \phi \right],
\ee
where $ \delta\lambda, \delta\blam, \delta\phi$ are defined in \p{N4}.

Let us note that the coefficients $(b,{\bar b},c)$ are defined up the to following gauge transformation:
\be\label{gauge}
\delta b = v_\lambda,\quad \delta {\bar b} = v_\blam, \quad \delta c = v_\phi \quad \Rightarrow \quad \delta B = {\dot v},
\ee
where $v$ is an arbitrary function of $(\phi, \lambda,\blam)$. Clearly, the new auxiliary component $\tilde B = B+{\dot v}$
will also transform  through a full time derivative. Just this freedom is reflected in the equations for
$(b,{\bar b},c)$ which are invariant under \p{gauge}.

In principle, the equations \p{B}, \p{eq}, \p{eq1} give a complete solution for our problem. But, when working in one
dimensional space, one may drastically simplify these expressions. Let us carefully explain all steps in this simplification.

First of all our basic Ansatz  is too general for one dimension. Indeed, without loss of generality one may write
\be\label{c1}
c=\partial_\phi {\tilde c} \quad \Rightarrow \quad c\dot\phi \rightarrow \frac{d}{dt}{\tilde c} - {\tilde c}_\lambda \dot\lambda -
{\tilde c}_\blam\dot\blam .
\ee
Obviously, one may discard the full time derivative $\frac{d}{dt}{\tilde c}$, which does not affect the transformation properties of $B$
through the full time derivative.
Moreover one may cancel two additional terms in \p{c1} by a proper redefinition of $b$ and ${\bar b}$. Therefore from the beginning
one may set $c=0$. This choice will restrict the gauge freedom \p{gauge} till residual transformations with a function ${\tilde v}(\lambda,\blam)$
depending on $(\lambda,\blam)$ only.

Next, introducing the new function $h$ as
\be\label{h}
h_\phi \equiv f
\ee
one may integrate last two equations in \p{eq}
\be\label{eqqq}
b_\phi=2ih_{\phi\lambda}+4\blam\frac{h_{\phi\phi}}{1+\lambda\blam},\;
{\bar b}_\phi=-2ih_{\phi\blam}+4\lambda\frac{h_{\phi\phi}}{1+\lambda\blam}
\ee
 to get
\be\label{bb}
b=2ih_\lambda + 4\blam \frac{h_\phi}{1+\lambda\blam}+{\hat b}(\lambda,\blam), \quad {\bar b}=-2ih_\blam + 4\lambda \frac{h_\phi}{1+\lambda\blam}
+{\overline{\hat b}}(\lambda,\blam).
\ee
Finally, representing ${\hat b}(\lambda,\blam)$ and ${\overline{\hat b}}(\lambda,\blam)$ as
\be
{\hat b}(\lambda,\blam) =\partial_\lambda b_0 +i \partial_\lambda b_1, \quad
{\overline{\hat b}}(\lambda,\blam) =\partial_\blam b_0 -i \partial_\blam b_1
\ee
one may see that the terms with $b_1$ may be absorbed into a new function $h$ by a redefinition of $h\rightarrow h+ b_1$, while the
terms with $b_0$ are just gauge transformations \p{gauge}, and we may choose the gauge $b_0=0$.

Plugging these $(b, {\bar b})$ \p{bb}  (with ${\hat b}(\lambda,\blam)={\overline{\hat b}}(\lambda,\blam)=0$) into the last equation in \p{eq} for them
\be\label{eqqqq}
b_\blam-{\bar b}_\lambda =-4i\frac{h_{\phi\phi}}{1+\lambda\blam}
\ee
we will get
\be\label{eq2}
h_{\phi\phi}+(1+\lambda\blam) h_{\lambda\blam}+i\lambda h_{\lambda\phi}-i\blam h_{\blam\phi}=0.
\ee
Thus, we have the following solution for the most general auxiliary
component $B$ which transforms through a full time derivative under $N=4$ supersymmetry:
\be\label{BB}
B=h_\phi  A +b\dot\lambda +{\bar b} \dot\blam + a \left(\bpsi{}^1\psi_1 -\bpsi{}^2\psi_2\right)+
   a_1 \bpsi{}^2\psi_2+a_2\bpsi{}^1\psi_1,
\ee
where
\bea\label{eq0}
&& a= -8\frac{h_{\phi\phi}}{1+\lambda\blam},\; a_1=-8ih_{\phi\blam}+8 \lambda\frac{h_{\phi\phi}}{1+\lambda\blam},\;
a_2=8ih_{\phi\lambda}+8 \blam\frac{h_{\phi\phi}}{1+\lambda\blam}, \nn
&&b=2ih_\lambda + 4\blam \frac{h_\phi}{1+\lambda\blam}, \quad {\bar b}=-2ih_\blam + 4\lambda \frac{h_\phi}{1+\lambda\blam}
\eea
and $h$ obeys the Laplace equation on $S^3$ \p{eq2}.

Now, with the newly defined auxiliary field $B$ we may add to the action \p{action2} a new generalized Fayet-Iliopoulos term
\be\label{FI2}
{\hat S} = S+ m\int dt B.
\ee
After excluding the auxiliary field $A$ we will have the following interaction terms in the components action
(as usual, we consider only the bosonic part of the action):
\be\label{pot2}
{\hat S}_{pot}=\int dt\left[ -\frac{m^2h_\phi^2}{g(1+\lambda\blam)}+2im\left(h_\lambda\dot\lambda -h_\blam\dot\blam\right)+
2m h_\phi \frac{\partial_t (\lambda\blam)}{1+\lambda\blam} \right].
\ee
Thus, we see that, with the newly defined Fayet-Iliopoulos term, we succeeded in constructing the most general action which
describes interactions with electric and magnetic fields and depends on the function $h$ which obeys equation \p{eq2}.
Let us observe that even for the $SU(2)\times U(1)$ sigma-model, which corresponds to the choice $g=1/(1+\lambda\blam)$, we have
a function freedom in the interaction terms. It will be interesting to analyze the possible self-interactions by choosing
a proper solution of the Laplace equation \p{eq2}.

Another trick we may to do is to dualize the auxiliary field $B$ instead of $A$, as follows:
\be\label{nB}
B={\dot u} \quad \Rightarrow \quad
{\tilde A} =\frac{1}{h_\phi} \left[  {\dot u} -b\dot\lambda -{\bar b} \dot\blam - a \left(\bpsi{}^1\psi_1 -\bpsi{}^2\psi_2\right)-
   a_1 \bpsi{}^2\psi_1-a_2\bpsi{}^1\psi_2\right].
\ee
Plugging the expression for $A$ \p{nB} into the action \p{action2}, we get a four dimensional sigma-model action
with the following bosonic part:
\be\label{finS}
{\hat S}=\int dt\; g(1+\lambda\blam)\left[ \left( \dot\phi +i \frac{ \dot\lambda \blam -\lambda \dot\blam }{1+\lambda\blam}\right)^2+
\frac{4 \dot\lambda\dot\blam}{(1+\lambda\blam)^2} +\frac{1}{4h_\phi^2}\left( {\dot u} -b\dot\lambda -{\bar b} \dot\blam+
2 h_\phi \frac{\partial_t (\lambda\blam)}{1+\lambda\blam}\right)^2 \right].
\ee
The action \p{finS} depends on two functions: the arbitrary metric $g(\phi, \lambda,\blam)$ and the auxiliary function $h(\phi, \lambda,\blam)$,
which obeys the Laplace equation on $S^3$. Thus, $N=4$ supersymmetry in $d=1$ leaves a lot of freedom in the sigma-model action.
It is interesting that if we choose
\be\label{papa}
g=2\frac{h_\phi}{1+\lambda\blam}
\ee
the action \p{finS} will exhibit the same target space geometry   which appears in the heterotic $(4,0)$ sigma-model
in $d=2$ \cite{heterotic}. The only simplification is that in $d=1$ one may completely solve the equations \p{eq} and
write explicitly the action in terms of the harmonic function $h$ only.

\setcounter{equation}0
\section*{Conclusion}
In this paper we considered the most general action for the $N=4, d=1$ nonlinear supermultiplet. We explicitly
constructed the most general interaction terms which depend on an arbitrary function $h$ obeying the Laplace
equation on $S^3$. Of course, in $d=1$ we have more freedom in the action, as compared to higher dimensions.
This freedom is reflected in the arbitrary metric $g$ which appear in the action. In order to get these results
we found the most general bosonic field $B$ which depends on the components of nonlinear supermultiplet and
transforms as a full time derivative under $N=4$ supersymmetry. The most general interaction is generated just
by a generalized Fayet-Iliopoulos terms build from this auxiliary component.

Being transformed through a full time derivative under $N=4, d=1$ supersymmetry, this auxiliary component $B$
may be dualized into a fourth scalar field. In a such way we constructed a four dimensional $N=4, d=1$ sigma-model
which possesses as a basic three-dimensional manifold the $SU(2)$ sigma-model. We analyzed the four-dimensional geometry in
the bosonic sector and found that it is not a hyper-K\"ahler one. With a particular choice of the metric $g$ \p{papa},
the geometry in the bosonic sector coincides with the one which appears in the heterotic $(4,0)$ sigma-model in $d=2$.

One of the interesting problems for further studying is to explicitly construct and analyze the potential terms, corresponding
to some interesting specific solution of the Laplace equation obeyed by the function $h$. We expect that some cases will
correspond to integrable systems, in full analogy with $N=4, d=1$ hyper-K\"ahler sigma models.
Another intriguing question concerns the $N=8, d=1$ nonlinear supermultiplet which is a direct dimensional reduction
of $N=2, d=4$ nonlinear supermultiplet \cite{{nlin},{klr},{HKLR}}. It has been known for a long time that starting from $N=2, d=4$
nonlinear supermultiplet in $N=1, d=4$ superspace on may dualize the scalar supermultiplet into a chiral one \cite{klr}.
The resulting metric of the new four-dimensional manifold will be the hyper-K\"ahler one. So, the same should be also true
for $N=8,d=1$ nonlinear supermultiplet. But at the same time, the action for  $N=8, d=1$ nonlinear supermultiplet
should admit a reduction to the $N=4, d=1$ case, which as we see does not support hyper-K\"ahler geometry. We hope to
resolve this paradox in a future publication.

Finally, we would like to stress that one dimensional extended supersymmetry is much simpler then the four-dimensional
one. So, the analysis of the possible sigma-model geometries is greatly simplified as compared to higher
dimensional cases. Indeed, the search for a generalized auxiliary component of the supermultiplet which may be
then dualized into an additional scalar component is equivalent in a some sense to performing a Legendre transform in $d=4$ \cite{HKLR},
being much simpler, though. Moreover, all $N=4$ and most of $N=8, d=1$ supermultiplets are off shell, what also
simplified the analysis. Unfortunately, the dualization of the auxiliary component will always produce a manifold with
an isometry. Nevertheless, the dualizations procedure can be generalized to give manifolds without
any obvious isometry. We hope to report the corresponding results elsewhere.

\section*{Acknowledgements}

The authors are grateful to A.~Shcherbakov and A.~Beylin for  useful discussions.
S.K. would like to thank the INFN--Laboratori Nazionali di Frascati for the
warm hospitality extended to him during the course of this work.

This work was partially supported by the European Community's Marie Curie Research Training Network under contract
MRTN-CT-2004-005104 Forces Universe, by INTAS under contract 05-7928 and by grants RFBR-06-02-16684, DFG~436 Rus~113/669/03.

\end{document}